\documentclass{ws-mpla}

\begin{document}

\markboth{H. Athar}
{Atmospheric and galactic tau neutrinos}

%
\catchline{}{}{}{}{}
%

\title{Atmospheric and galactic tau neutrinos\footnote{Invited 
talk given at 2003 International Symposium on Cosmology and Particle
Astrophysics (CosPA 2003), Nov. 13$-$15, 2003, Taipei, Taiwan.}}

\author{\footnotesize H. ATHAR}

\address{Physics Division, National Center for Theoretical
Sciences,\\
101 Section 2, Kuang Fu Road, Hsinchu 300, Taiwan\\
athar@phys.cts.nthu.edu.tw}

\maketitle

\pub{Received (Day Month Year)}{{\it In memoriam of Akbar Husain}}

\begin{abstract}
Neutrinos with energy greater than GeV are copiously produced
in the $p(A,p)$ interactions occurring in the earth atmosphere and in
our galactic plane. A comparison of the tau and mu neutrino 
flux in the presence of neutrino oscillations from
these two astrophysical sites is presented. It is pointed out 
that the galactic plane tau neutrino flux dominates over the
 downward going atmospheric tau neutrino flux at much lower
energy value than that for the dominance of the mu
neutrino flux from these two sites. Future prospects for
possible observations of galactic tau neutrino flux are also briefly mentioned.

\end{abstract}

\ccode{PACS Nos.: 98.38-j,13.85.Tp,14.60.Pq}
\section{Introduction}	
A present day main motivation for the extra-terrestrial neutrino astronomy 
is to obtain first evidence of tau neutrinos from the cosmos around us 
 above the relatively 
well known atmospheric neutrino background.\cite{Athar:2002rr} The 
tau neutrinos are an unavoidable consequence of neutrino flavor
mixing as suggested by the high statistics Super-Kamiokande detector 
(SKK).\cite{Fukuda:2000np}

A recent SKK analysis of the 
atmospheric neutrino data imply the following range of neutrino mixing parameters   
\begin{equation}
 \delta m^{2}=(1.3-3.0)\cdot 10^{-3}\, \, \, {\rm eV}^{2}, \, \, \, 
 \sin^{2}2\theta >0.9.
\label{range}
\end{equation}
This is a  $90\% \, {\rm C.L.}$ range 
with the best fit values approximately given by 
$\delta m^{2}=2\cdot 10^{-3}\, \, \, {\rm eV}^{2}$ and  $\sin^{2}2\theta =1$
 respectively. 
 This range of neutrino mixing parameters results in purely 
two flavor oscillation explanation of 
the zenith angle dependence  of the atmospheric 
mu neutrino deficit, along with another indication. 
 The tau neutrinos as a result of these 
 $\nu_{\mu}\to \nu_{\tau}$ oscillations are so far however identified on 
 statistical basis only (rather than on event by event basis).\cite{icrr2003}  
 On the other hand, the total number of 
 observed non tau neutrinos are by now greater than $10^{4}$ from various
 detectors ranging in energy between $10^{-1}$ GeV to $10^{3}$ GeV.\cite{Kajita:2000mr}
 Thus, it is of some interest to estimate the tau neutrino flux from the 
earth atmosphere as well as from the nearby astrophysical sites 
 such as our galactic plane to provide a 
more complete basis for the hypothesis of $\nu_{\mu}\to \nu_{\tau}$
 oscillations. 

 The neutrino oscillation probability in the two neutrino flavor
approximation is  
\begin{equation}
 P(\nu_{\mu}\to \nu_{\tau})=\sin^{2}2\theta \sin^{2}
 \left(\frac{l}{l_{\rm osc}}\right).
\label{osc-prob}
\end{equation}
Here $l$ can be estimated using  
\begin{equation}
 l=\sqrt{(R_{\oplus}+h)^{2}-R^{2}_{\oplus}\sin^{2}\phi}-R_{\oplus} \cos \phi.
\label{len}
\end{equation}
The $l$ is the distance between the detector and the height  
 at which the atmospheric mu neutrinos are produced. 
 The $R_{\oplus}\simeq 6.4 \cdot 10^{3}$ km is the earth radius, and  $h=15$ km  
 is the mean altitude at which the atmospheric mu neutrinos are produced.
 In general, $h$ is not only a function of the zenith angle $\phi$, the neutrino flavor
 but also the neutrino energy.\cite{Gaisser:1997eu} Also
\begin{equation}
 \frac{1}{l_{\rm osc}}=
 \frac{1.27}{{\rm km}} \left(\frac{{\rm GeV}}{E}\right) 
 \left( \frac{\delta m^{2}}{{\rm eV^{2}}} \right),
\label{osc-len}
\end{equation}
in usual notation. 

This paper is organized as follows. In section 2, the mu and tau neutrino flux 
originating from the earth atmosphere and the 
galactic plane is briefly discussed. In section 3, the neutrino oscillation effects 
are studied for both. In section 4, the limited future prospects 
for possible observations of galactic tau neutrinos are mentioned, 
whereas in section 4, conclusions are presented. 
\section{Atmospheric and galactic neutrino flux}
Briefly, the incoming cosmic rays interact with the air nuclei $A$, in the 
 earth atmosphere and give rise to mu neutrino flux. 
For  $1 \leq E/{\rm GeV} \leq 10^{3}$, the $\pi^{\pm}$, $K$ production and 
 their direct and indirect decays
 are the main sources of mu neutrinos, both being in region of conventional 
  mu neutrino production.\cite{1962} The absolute normalization
of the conventional atmospheric neutrino flux is presently known to be no better
than (20$-$25)\%.\cite{Battistoni:1999at}

For present estimates, the mu neutrino flux is taken from 
 Ref.~\refcite{Honda:1995hz}. These are 
 neutrino flux calculations in one dimension without geomagnetic field
 effects. The up down
mu neutrino flux is taken to be the same, as the present discussion is
independent of any specific detector. At higher energy, the prompt mu
neutrino production from $D$'s dominates over the conventional one.\cite{Volkova:gh}

\begin{figure}[th]
\centerline{\psfig{file=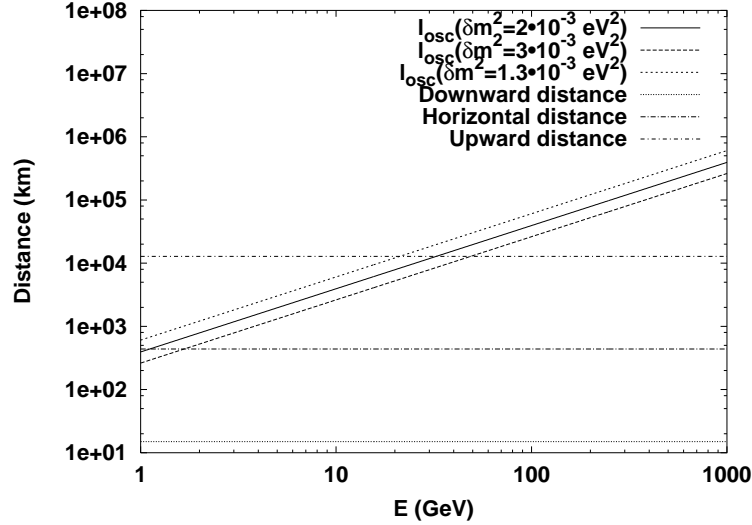,width=4.0in}}
\vspace*{8pt}
\caption{The oscillation length ($l_{\rm osc}$) in km as a function of neutrino 
 energy in GeV. The three general distances 
 traversed by mu neutrinos in earth atmosphere are also shown
 (the horizontal lines). See text for more details.}
\label{figone}
\end{figure}

The atmospheric tau neutrino 
flux arises mainly from $D^{\pm}_{S}$  and is
calculated in Ref.~\refcite{Pasquali:1998xf,Athar:2001jw}. 
 The Quark Gluon String Model (QGSM) is used in Ref.~\refcite{Athar:2001jw} to
 model the $pA$ interactions.  The low energy 
atmospheric tau neutrino flux is essentially isotropic.\cite{Pasquali:1998xf} 
 For $E\leq  10^{3}$ GeV, the atmospheric tau neutrino flux is obtained by 
 re scaling w.r.t new cosmic ray flux
spectrum, taking  it to be dominantly the protons.\cite{Gaisser:2002jj,ICRC2001}

The galactic mu neutrino flux for $E \geq 10^{2}$ GeV 
 is calculated in Ref.~\refcite{Ingelman:1996md}, whereas 
the galactic plane tau neutrino flux is calculated in Ref.~\refcite{Athar:2001jw}.
 These calculations consider $pp$ interactions inside the galaxy with 
 target proton number density $\sim $ 1/cm$^{3}$ along the galactic plane. 
The tau neutrino production is rather suppressed in the
galactic plane relative to mu neutrino production.

 The galactic plane mu and tau neutrino flux for $E\leq  10^{3}$
 GeV is obtained by re scaling w.r.t new cosmic ray flux
spectrum. The distance inside the galactic
plane is taken to be $\sim 10 $ kpc where 1 pc $\simeq 3\cdot 10^{13}$ km.  

The  mu neutrino flux is larger than the tau 
neutrino flux for $E\leq  10^{3}$ GeV from the two sites. 
A detailed study that explicitly estimates the tau 
neutrino flux from the two sites for low energy indicates that the 
simple re scaling adopted here is a good approximation for $E \geq 10$ GeV.\cite{us} 
\section{Effects of neutrino oscillations}
In the two flavor approximation, the total tau neutrino flux is

\begin{equation}
 \frac{{\rm d} N_{\nu_{\tau}}}{{\rm d(log_{10}}E)} =
 P(\nu_{\mu}\to \nu_{\tau})\cdot \frac{{\rm d}N^{0}_{\nu_{\mu}}}{{\rm d(log_{10}}E)} +
 P(\nu_{\mu}\to \nu_{\mu})\cdot \frac{{\rm d}N^{0}_{\nu_{\tau}}}{{\rm d(log_{10}}E)},
\label{tot}
\end{equation}
where $P(\nu_{\mu}\to \nu_{\tau})$ is given by Eq. (\ref{osc-prob})
 and $P(\nu_{\mu}\to \nu_{\mu})=1-P(\nu_{\mu}\to \nu_{\tau})$.  The 
\begin{figure}[th]
\centerline{\psfig{file=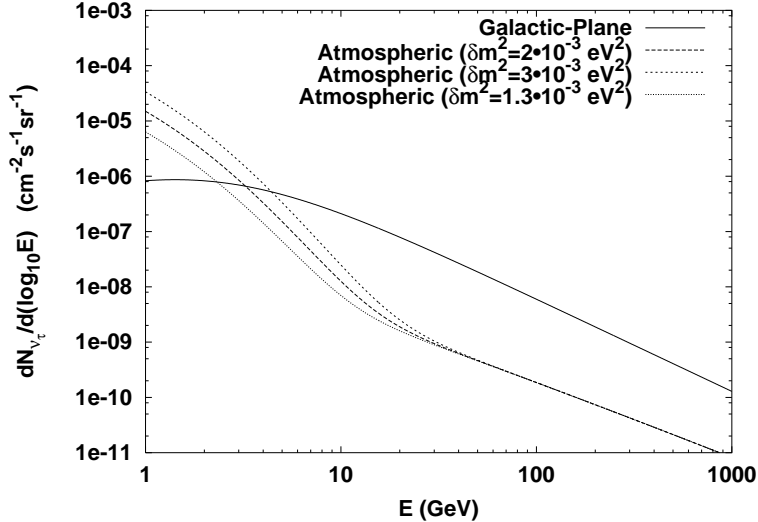,width=4.0in}}
\vspace*{8pt}
\caption{The comparison of the galactic plane and the downward going atmospheric tau
 neutrino flux in the presence of neutrino oscillations as a function of neutrino
 energy.}
\label{figtwo}
\end{figure}
${\rm d}N^{0}_{\nu}/{\rm d(log_{10}}E)$ is in units of cm$^{-2}$s$^{-1}$sr$^{-1}$ 
 and is taken according to discussion in section 2.

Three general directions in the earth atmosphere 
are considered as representative examples to compare the atmospheric tau neutrino
flux with the galactic one in the {\tt presence} of neutrino oscillations. 
 These are downward, horizontal
and upward. Fig. \ref{figone} depicts the $l_{\rm osc}$ given by 
Eq. (\ref{osc-len}) for the range of $\delta m^{2}$ given by Eq. (\ref{range}) with maximal 
 mixing.
 The three distances are taken from Eq. (\ref{len}), with, for instance,  the downward 
distance is obtained by setting $\phi =0$. The horizontal distance is 
obtained by setting $\phi = \pi/2$.

Using Eq. (\ref{tot}), the total {\tt downward} going atmospheric 
tau neutrino flux is estimated. It is then compared with the total galactic plane 
 tau neutrino flux in Fig. \ref{figtwo} for the whole
range of $\delta m^{2}$ with maximal mixing. The distance $l$ for galactic plane neutrinos
is taken as $\sim $ 5 kpc. Since 
$l_{\rm osc} \ll l$, the galactic plane mu neutrinos oscillate  
before reaching the earth. Also, note that this flux is averaged out
 for the whole range of $\delta m^{2}$ in the entire considered energy range. 
 The effect of different 
$\delta m^{2}$ values diminishes for $E \geq 50$ GeV for total 
 atmospheric tau neutrino flux. From the figure, it can be seen 
that the galactic plane tau neutrino flux starts {\tt dominating} over
the downward going atmospheric tau neutrino flux even for 
$E$ as low as 10 GeV in the presence of neutrino oscillations. 
 This is a very specific feature of {\tt tau neutrinos},
 and is absent for mu neutrinos. This specific 
behavior has to do with the {\tt neutrino oscillations}. The
galactic plane tau neutrino flux for $1\leq E/{\rm GeV} \leq 10^{3}$ in the
presence of neutrino oscillations can be parameterized as
\begin{equation}
 \frac{{\rm d} N_{\nu_{\tau}}}{{\rm d(log_{10}}E)} =
 1.31\cdot 10^{-5} \cdot \frac{E^{1.07}}{\left[ E+2.15\exp({-0.21\sqrt{E}}) \right]^{2.74}},
\label{parameterize}
\end{equation}
\begin{figure}[th]
\centerline{\psfig{file=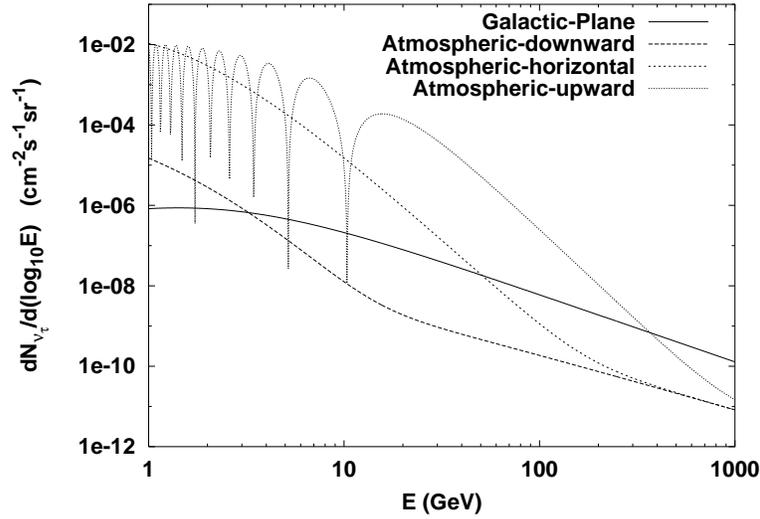,width=4.0in}}
\vspace*{8pt}
\caption{The galactic plane and the atmospheric tau
 neutrino flux in the presence of neutrino oscillations for 
 the three general directions in the earth atmosphere as a function of neutrino energy. 
 The neutrino mixing parameter values used here are the approximate best fit values,
 i.e., $\delta m^{2}=2\cdot 10^{-3}$ eV$^{2}$ and $\sin^{2}2\theta =1$.}
\label{figthree}
\end{figure}
where ${\rm d} N_{\nu_{\tau}}/{\rm d(log_{10}}E)$ is in units of 
cm$^{-2}$s$^{-1}$sr$^{-1}$ and on r.h.s. $E$ is in units of GeV.

In Fig. \ref{figthree}, the galactic plane tau neutrino 
flux is compared with the atmospheric tau neutrino flux,
 using Eq. (\ref{tot}) for the three general 
directions for the atmospheric tau neutrino flux reaching the detector. 
 Here the best fit values of the neutrino
mixing parameters are used.  The oscillatory nature of the upward 
going tau neutrino flux can be seen from Eq. (\ref{osc-prob}).
The cross over for the galactic tau neutrinos relative to the 
 horizontal atmospheric tau neutrinos 
occurs at $\sim $ 50 GeV, whereas the same occurs for the upward
direction at $\sim $ 400 GeV. The total atmospheric tau neutrino flux is maximum
in the upward direction (see Fig. \ref{figthree}). It is
minimum in downward direction, relative to the galactic plane 
tau neutrino flux in the presence of neutrino oscillations, 
 owing to the behavior of $l/l_{\rm osc}$ ratio as a function of neutrino energy.  
 Fig. \ref{figthree} indicates that zenith angle 
dependence of the total tau neutrino flux can at least in
principle help to distinguish between atmospheric and
 non atmospheric tau neutrino flux.  The galactic tau 
neutrino flux transverse to the galactic plane is 
three to four orders of magnitude smaller than the
galactic plane one.\cite{Athar:2001jw}
\begin{figure}[th]
\centerline{\psfig{file=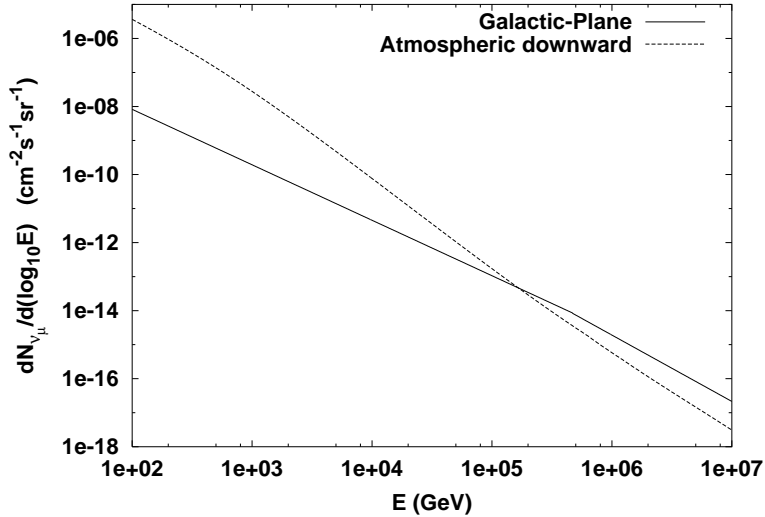,width=4.0in}}
\vspace*{8pt}
\caption{Comparison of the downward going atmospheric mu neutrino
 flux and the galactic plane mu neutrino flux in the 
 presence of neutrino oscillations. The galactic plane
 mu neutrino flux starts dominating over the atmospheric one 
 only for $E \geq 10^{5}$ GeV.}
\label{figfour}
\end{figure}

Fig. \ref{figfour} gives a comparison of the downward going atmospheric and
the galactic plane mu neutrino flux in the presence of neutrino 
 oscillations. For this comparison, mu neutrino 
 flux is taken from Ref.~\refcite{Gondolo:1995fq} without re scaling
 for $E \leq 10^{3}$ GeV. This 
mu neutrino flux includes contribution from the $D$'s  
 for $E \geq 6.3\cdot 10^{5}$ GeV. 
The total mu neutrino flux is
 estimated according to Eq. (\ref{tot}) with appropriate
 modifications for the best fit values of the two neutrino mixing parameters. 
 In contrast to the 
possibility of seeing the galactic plane with multi GeV 
tau neutrinos, note here that with mu neutrinos, it 
can occur only for $E \geq 10^{5}$ GeV.

A relevant remark is that for the best fit values of the neutrino mixing parameters, 
the $P(\nu_{\mu}\to \nu_{\tau})$ is relatively large along the 
horizontal and upward directions in the earth atmosphere 
 [see Eq. (\ref{osc-prob}) and Fig. \ref{figone}] for $1 \leq E/{\rm GeV} \leq 10$. 
 So essentially the atmospheric mu neutrino flux in the absence of 
neutrino oscillations alone determine the total atmospheric
tau neutrino flux in comparison with the total galactic
 tau neutrino flux.  
\section{Prospects for possible observations}
For $10 \leq E/{\rm GeV} \leq 10^{3}$, a signature for the tau neutrinos is to 
 measure the energy spectrum of the tau lepton induced electromagnetic and hadronic 
 showers 
produced in tau neutrino nucleon interactions
occurring inside a densely instrumented Cherenkov radiation detector.\cite{Stanev:1999ki} 
Though, it is a challenging task to distinguish between tau and non tau 
neutrinos for the present generation of detectors in the above 
 energy range\cite{Hall:1998ey}, however certain shower signatures remain
distinctive  for tau neutrinos.\cite{Stanev:1999ki}

The galactic tau neutrino induced shower production  rate can be 
 approximately estimated  by convolving the 
total galactic tau neutrino flux in the presence of neutrino oscillations, given by 
 Eq. (\ref{parameterize}) with the $\sigma^{\rm tot}_{\nu_{\tau}N}$,
where  
$\sigma^{\rm tot}_{\nu_{\tau}N}=\sigma^{\rm CC}_{\nu_{\tau}N}+
 \sigma^{\rm NC}_{\nu_{\tau}N}$ for $10 \leq E/{\rm GeV} \leq 10^{3}$. For recent evaluations of 
 $\sigma^{\rm tot}_{\nu_{\tau}N}$, see Ref.~\refcite{Paschos:2001np}.
 The possible tau lepton polarization effects\cite{Hagiwara:2003di}
 are not taken into account in the
 event rate estimates presented here.

\begin{table}[h]
\tbl{Galactic tau neutrino induced shower event rate. Details are given in the text.}
{\begin{tabular}{@{}cc@{}} \toprule
 Energy Bin  (GeV)   &  
$N_{\nu_{\tau}}({\rm Mt}\cdot {\rm yr} \cdot 2\pi {\rm \, sr})^{-1}$ \\   \colrule
 10  $-$ 31.62 &  0.52 \\
 31.62 $-$ 100   &  0.50 \\
 100 $-$  316.2 &  0.30 \\
 316.2 $-$  1000  & 0.17  \\ \botrule
\label{tableone}
\end{tabular}}
\end{table}

  Table \ref{tableone}
gives the galactic tau neutrino induced shower event rate for a 1 Mega ton detector,
 in units of $({\rm Mt}\cdot {\rm yr})^{-1}$
 in $2\pi $ steradians of upper hemi-sphere  
 in four wide logarithmically equally spaced energy bins.  The table indicates
 that with a 3 to 5 year data collection time for a one Mega ton 
detector, the galactic tau neutrino induced shower event rate can be in the range of 
(1$-$10) for $E \geq 10$ GeV.
 This detector faces only the downward going atmospheric tau neutrino
 flux as background to the dominant galactic plane tau neutrino 
 flux in the presence of neutrino oscillations.  

\section{Conclusions}
1. The effects of neutrino oscillations on 
low energy ($E \leq 10^{3}$ GeV) tau neutrino flux produced in the earth atmosphere  
 and in our galactic plane are presented in two neutrino flavor approximation.

2.  The galactic plane should be observable
with tau neutrinos with energy $\geq 10$ GeV, depending on the orientation 
of the concerned detector w.r.t. galactic center/plane at the time of observation. 
The observation of galactic plane with multi GeV {\tt tau neutrinos} is in sharp 
 contrast to the case
of mu neutrinos with which the galactic plane is observable only with energy 
$\geq  10^{5}$ GeV for the same orientation of the detector. 

3. This observation may also have some relevance
for the long baseline experiments searching for the tau neutrinos in
$\nu_{\mu}\to \nu_{\tau}$ oscillations.\cite{lbl}   
\section*{References}


\vspace*{6pt}

\begin{thebibliography}{0}
%
%
%
\bibitem{Athar:2002rr}
H.~Athar,
{\it Nucl.\ Phys.\ Proc.\ Suppl.\  }  {\bf 122}, 305 (2003).
 For a recent brief review article, see, {\em ibid}.,
arXiv:hep-ph/0308188 [to appear in {\it Chin. J. Phys.}].
%
%
%
\bibitem{Fukuda:2000np}
S.~Fukuda {\it et al.}  [Super-Kamiokande Collaboration],
{\it Phys.\ Rev.\ Lett.\ } {\bf 85}, 3999 (2000).
%
%
%
\bibitem{icrr2003}
 See, for instance, plenary talk given by Y. Suzuki at 
 {\it 28th Intl. Cosmic Ray Conf. }(ICRC 2003), 
 Tokyo, Japan, 31 July $-$ 07 August, 2003 
 (http://www-rccn.icrr.u-tokyo.ac.jp/icrc2003/talks/Plenary-Suzuki.pdf, 
to appear in its proceedings).
%
%
%
\bibitem{Kajita:2000mr}
T.~Kajita and Y.~Totsuka,
{\it Rev.\ Mod.\ Phys.\ } {\bf 73}, 85 (2001).
 For a recent discussion of up coming astrophysical neutrino 
 telescopes, see, 
A.~B.~McDonald {\it et al}., 
arXiv:astro-ph/0311343.
%
%
%
\bibitem{Gaisser:1997eu}
T.~K.~Gaisser and T.~Stanev,
{\it Phys.\ Rev.\ } {\bf D57}, 1977 (1998);
 {\em ibid.,}
{\it Nucl.\ Phys.\ Proc.\ Suppl.\ } {\bf 70}, 335 (1999);
 M.~Honda, T.~Kajita, K.~Kasahara and S.~Midorikawa,
 {\it Phys.\ Rev.\ } {\bf D64}, 053011 (2001).
%
%
%
\bibitem{1962}
G. T. Zatsepin and V. A. Kuzmin, 
{\it Sov. Phys. JETP} {\bf 14}, 1294 (1962);
L.~V.~Volkova,
{\it Sov.\ J.\ Nucl.\ Phys.\ } {\bf 31}, 784 (1980);
T. K. Gaisser and Todor Stanev, S. A. Bludman and H. Lee,
{\it Phys.\ Rev.\ Lett.\  }{\bf 51}, 223 (1983);
A.~Dar,
{\it Phys.\ Rev.\ Lett.\  }{\bf 51}, 227 (1983);
K.~Mitsui, H.~Komori and Y.~Minorikawa,
{\it Nuovo Cim.\ } {\bf C9}, 995 (1986);
E.~V.~Bugaev and V.~A.~Naumov,
{\it Sov.\ J.\ Nucl.\ Phys.\  }{\bf 45}, 857 (1987);
T.~K.~Gaisser, T.~Stanev and G.~Barr,
{\it Phys.\ Rev.\  }{\bf D38}, 85 (1988);
A.~V.~Butkevich, L.~G.~Dedenko and I.~M.~Zheleznykh,
{\it Sov.\ J.\ Nucl.\ Phys.\  }{\bf 50}, 90 (1989);
E.~V.~Bugaev and V.~A.~Naumov,
{\it Phys.\ Lett.\  }{\bf B232}, 391 (1989);
G.~Barr, T.~K.~Gaisser and T.~Stanev,
{\it Phys.\ Rev.\  }{\bf D39}, 3532 (1989);
H. Lee and Y. Koh, {\it Nouvo Cim.}  {\bf B105}, 884 (1990);
M.~Honda, K.~Kasahara, K.~Hidaka and S.~Midorikawa,
{\it Phys.\ Lett.\  }{\bf B248}, 193 (1990);
P.~Lipari,
{\it Astropart.\ Phys.\  }{\bf 1}, 195 (1993);
D.~H.~Perkins,
{\it Astropart.\ Phys.\  }{\bf 2}, 249 (1994);
T.~K.~Gaisser, {\it et al}., 
{\it Phys.\ Rev.\  }{\bf D54}, 5578 (1996);
M.~Honda, T.~Kajita, K.~Kasahara and S.~Midorikawa,
{\it Prog.\ Theor.\ Phys.\ Suppl.\  }{\bf 123}, 483 (1996);
P.~Lipari, T.~Stanev and T.~K.~Gaisser,
{\it Phys.\ Rev.\  }{\bf D58}, 073003 (1998).
%
%
%
\bibitem{Battistoni:1999at}
 See, for instance, 
 G.~Battistoni {\it et al}., 
{\it Astropart.\ Phys.\ } {\bf 12}, 315 (2000).
 For a recent semi-analytic treatment, see, T.~K.~Gaisser,
{\it Astropart.\ Phys.\ } {\bf 16}, 285 (2002).
%
%
%
\bibitem{Honda:1995hz}
M.~Honda, T.~Kajita, K.~Kasahara and S.~Midorikawa,
{\it Phys.\ Rev.\ } {\bf D52}, 4985 (1995);
 V.~Agrawal, T.~K.~Gaisser, P.~Lipari and T.~Stanev,
 {\it Phys.\ Rev.\   } {\bf D53}, 1314 (1996).
%
%
%
\bibitem{Volkova:gh}
H.~Inazawa and K.~Kobayakawa,
{\it Prog.\ Theor.\ Phys.\ } {\bf 69}, 1195 (1983);
L.~V.~Volkova and G.~T.~Zatsepin,
{\it Sov.\ J.\ Nucl.\ Phys.\  }{\bf 37}, 212 (1983);
C.~Castagnoli {\it et al.}, 
{\it Nuovo Cim.\  }{\bf A82}, 78 (1984);
H.~Inazawa, K.~Kobayakawa and T.~Kitamura,
{\it J.\ Phys.\  }{\bf G12}, 59 (1986);
L.~V.~Volkova, W.~Fulgione, P.~Galeotti and O.~Saavedra,
{\it Nuovo Cim.\ } {\bf C10}, 465 (1987);
E.~Zas, F.~Halzen and R.~A.~Vazquez,
{\it Astropart.\ Phys.\  }{\bf 1}, 297 (1993);
G.~Battistoni {\it et al}., 
{\it Astropart.\ Phys.\  }{\bf 4}, 351 (1996);
V.~A.~Naumov, T.~S.~Sinegovskaya and S.~I.~Sinegovsky,
{\it Nuovo Cim.\  }{\bf A111}, 129 (1998);
E.~V.~Bugaev {\it et al}., 
{\it Phys.\ Rev.\ } {\bf D58}, 054001 (1998);
L.~Pasquali, M.~H.~Reno and I.~Sarcevic,
{\it Phys.\ Rev.\  }{\bf D59}, 034020 (1999);
G.~Gelmini, P.~Gondolo and G.~Varieschi,
{\it Phys.\ Rev.\  }{\bf D61}, 036005 (2000);
L.~V.~Volkova and G.~T.~Zatsepin,
{\it Phys.\ Atom.\ Nucl.\  }{\bf 64}, 266 (2001);
C.~G.~S.~Costa, F.~Halzen and C.~Salles,
{\it Phys.\ Rev.\  }{\bf D66}, 113002 (2002);
A.~M.~Stasto,
arXiv:astro-ph/0310636;
 H.~Athar, K.~Cheung, G.~L.~Lin and J.~J.~Tseng,
arXiv:astro-ph/0311586. 
%
%
%
\bibitem{Pasquali:1998xf}
L.~Pasquali and M.~H.~Reno,
{\it Phys.\ Rev.\ } {\bf D59}, 093003 (1999).
%
%
%
\bibitem{Athar:2001jw}
H.~Athar, K.~Cheung, G.~L.~Lin and J.~J.~Tseng,
{\it Astropart.\ Phys.\ } {\bf 18}, 581 (2003).
%
%
%
\bibitem{Gaisser:2002jj}
T.~K.~Gaisser and M.~Honda,
 {\it Ann.\ Rev.\ Nucl.\ Part.\ Sci.\  }{\bf 52}, 153 (2002).
%
%
%
\bibitem{ICRC2001}
T. K. Gaisser, M. Honda, P. Lipari, and T. Stanev, 
 talk given at {\it 27th Intl. Cosmic Ray Conf. }(ICRC 2001), Hamburg, Germany,
07$-$15 August, 2001, pp. 1643-1646.
%
%
%
\bibitem{Ingelman:1996md}
For a recent calculation, see, G.~Ingelman and M.~Thunman,
arXiv:hep-ph/9604286.
%
%
%
\bibitem{us}
 H. Athar, Fei-Fain Lee, Guey-Lin Lin, to be submitted. 
%
%
%
\bibitem{Gondolo:1995fq}
 M.~Thunman, G.~Ingelman and P.~Gondolo,
{\it Astropart.\ Phys.\ } {\bf 5}, 309 (1996).
%
%
%
\bibitem{Stanev:1999ki}
T.~Stanev,
{\it Phys.\ Rev.\ Lett.\ } {\bf 83}, 5427 (1999).
%
%
%
\bibitem{Hall:1998ey}
L.~J.~Hall and H.~Murayama,
{\it Phys.\ Lett.\ } {\bf B463}, 241 (1999).
 See also, F.~Vissani and A.~Yu.~Smirnov,
{\it Phys.\ Lett.\ } {\bf B432}, 376 (1998).
%
%
%
\bibitem{Paschos:2001np}
E.~A.~Paschos and J.~Y.~Yu,
{\it Phys.\ Rev.\ } {\bf D65}, 033002 (2002);
 S.~Kretzer and M.~H.~Reno,
{\it Phys.\ Rev.\ } {\bf D66}, 113007 (2002);
 {\em ibid}., 
arXiv:hep-ph/0306307.
%
%
%
\bibitem{Hagiwara:2003di}
K.~Hagiwara, K.~Mawatari and H.~Yokoya,
 {\it Nucl.\ Phys.\ } {\bf B668}, 364 (2003).
%
%
%
\bibitem{lbl}
See, for instance, http://www.hep.anl.gov/ndk/longbnews/ for details of the 
 long base line neutrino experiments.
%
%
%
\end{thebibliography}
\end{document}